\documentclass[prb,preprint,superscriptaddress,twocolumn,10pt]{revtex4-1} 

\usepackage{amsmath}  
\usepackage{amsfonts} 
\usepackage{graphicx} 
\usepackage{esvect}
\usepackage{mathtools}
\usepackage{subcaption}
\usepackage{xcolor}
\usepackage{appendix}

\usepackage{siunitx} 
\usepackage{placeins}

\usepackage{float}

\usepackage{pbox}

\usepackage{url}

\usepackage{xcolor}
\pagecolor{white}

\begin{document}


\title{Edge of Infinity: The Clash between Edge Effect and Infinity Assumption \\ for the Distribution of Charge on a Conducting Plate}

%
%
%
%
%

\author{Quy C. Tran} 
\affiliation{High School for Gifted Students, Vietnam National University, Hanoi 100000, Vietnam.}

\author{Nam H. Nguyen}
\affiliation{Duke Univeristy, Durham, NC 27708, USA }

\author{Thach A. Nguyen}
\affiliation{Le Hong Phong High School for the Gifted, Ho Chi Minh 700000, Vietnam}

\author{Trung V. Phan}
\email{trung.phan@yale.edu}
\affiliation{Department of Molecular, Cellular, and Developmental Biology, Yale University, New Haven, CT 06520, USA}

\begin{abstract}
We re-examine a familiar problem given in introductory to physics courses, about determining the induce charge distribution on an uncharged ``infinitely-large'' conducting plate when placing parallel to it a uniform charged nonconducting plate of the same size. We show that, no matter how large the plates are, the edge effect will always be strong enough to influence the charge distribution deep in the central region, which totally destroyed the infinity assumption (that the surface charge densities on the two sides are uniform and of opposite magnitudes). For a more detail analysis, we solve the Poisson's equation for a similar setting in two-dimensional space and obtain the exact charge distribution, helping us to understand what happen how charge distributes at the central, the asymptotic and the edge regions.
\end{abstract}

\maketitle

\date{\today}

\section{A Curious Puzzle}

\ \ 

One of the authors has been teaching in college for many years, and during that time there is always the following homework problem (or similar) for the introductory course to electromagnetic every year:

\ \

\begin{figure}[!htb]
\centering
\includegraphics[width=0.45\textwidth]{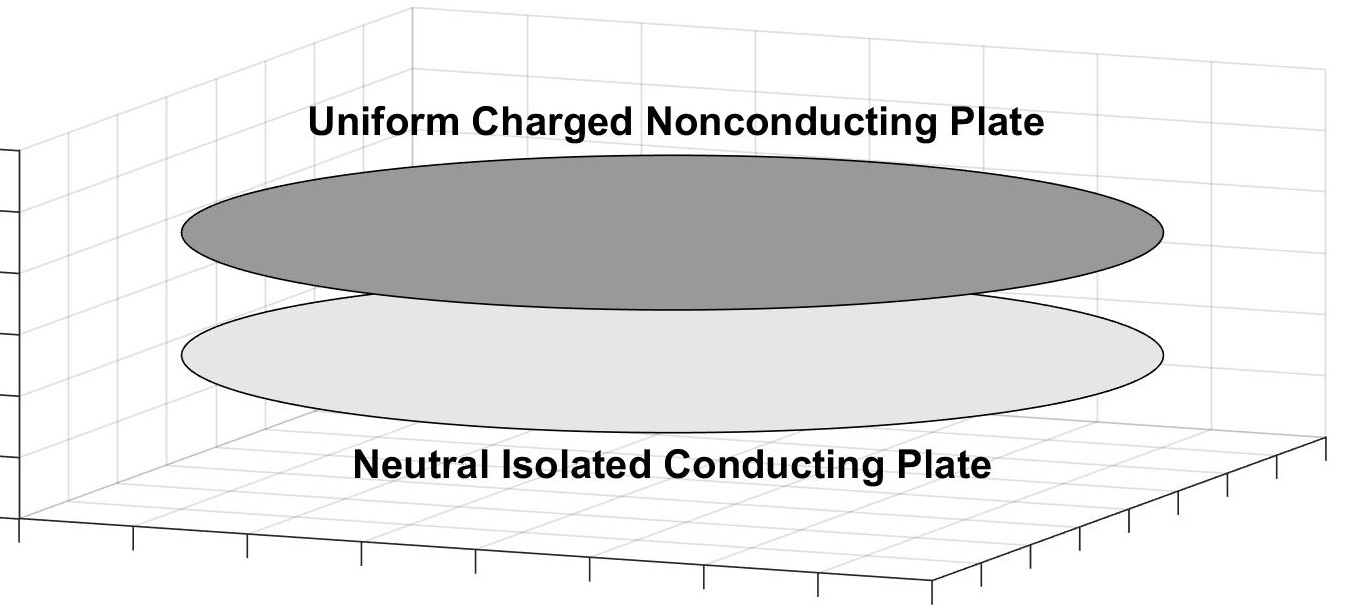}
\caption{A physical setting for the problem of interests in three-dimensional space, with the separation $H$ between these parallel plates are much smaller than their size $R$.}
\label{fig:3D_Setting}
\end{figure}

{\it ``An infinite nonconducting plate with uniform surface charge density $+\sigma_0^*$ is placed in parallel to an uncharged conducting plate. Find the induced charge distribution on both sides of the conducting plate.''} See Fig. \ref{fig:3D_Setting}.

\ \ 

\ \ 

The official solution to this problem, either given by the staffs running the course or handed down through out the years, is to use the infinity assumption so that the charge density on the side closer to the nonconducting plate $\sigma_\uparrow$ and on the side further to the nonconducting plate $\sigma_\downarrow$ are uniform and of equal magnitude but opposite signs $\sigma_\uparrow = -\sigma_\downarrow$ (for total charge neutrality). Then, from the condition that there is no electrical field $\vec{E}$ inside the conducting plate, the following relation has to be satisfied:
\begin{equation}
\left|\vec{E}\right|=\frac{\sigma_0^*}{2\epsilon} + \frac{\sigma_\uparrow}{2\epsilon} - \frac{\sigma_\downarrow}{2\epsilon}= 0 \ \Rightarrow \ \sigma_\uparrow = -\sigma_\downarrow = -\frac{\sigma_0^*}{2} \ .
\label{3D_wrong_solution}
\end{equation}
In other words, it is estimated that the surface densities on both side of the uncharged conducting plate with have the same magnitude of half the surface density on the nonconducting plate.

\ \ 

Let us take a step back and try to understand what is so puzzling about this solution, even though it might seems perfectly sounded at first. There is no such thing as infinite plates -- only an ``infinitely-large'' ones can exist. Say, the (radial) size of the plates in the problem is $R$ and the separated distance between them is $H$, then the infinity assumption can be provoked when considering what happen deep in the central region when $R\gg H$ and. It is physical to say $R \rightarrow \infty$, but it is unphysical to say $R = \infty$. In other words, there must be a boundary, an edge to this infinity. While the edge effect typically contributes the most to the fringe electrical field far away from the central region, it can influence the charge distribution very drastically. Consider a circular conducting plate of radius $R$ having total charged $Q$, what is the surface charge density $\sigma(r)$ (on both sides) at radial position $r$ from the center of the plate? This is a famous question, one of the few in classical physics which can be answered straight-forwardly by adding extra-dimensions \cite{phan2021curious}. J. J. Thomson has given an elegant geometric argument for the charge distribution on a plate as the limiting case of an oblate ellipsoid \cite{kelvin1884reprint}:
\begin{equation}
\sigma(r) = \frac{Q}{4\pi R^2} \bigg( 1 - \frac{r}{R} \bigg)^{-1/2} \ ,
\label{3D_circular_disk_charge_distritution}
\end{equation}
which is also equal to the projected surface charge distribution on the sphere onto its equator's plane. For a total charge that scales with the plate's area, i.e. $Q \propto R^2$, at the central region the charge distribution $\sigma(0)$ can be a substantial finite value. Such situation can happen (and indeed does happen) for the homework problem of interests.

\ \ 

We will now show some evidences for that. We can separate the main problem into two: find the charge distribution for (i) the same setting but with the conducting plate grounded, and (ii) a charged isolated conducting plate. Then, by superpose these two problems (and make sure that the total induced charge is equal to zero), we arrive at the answer to the original. Problem (i) is similar to a capacitor in which one of the plate is connected to the ground and the other is kept at a constant electrical potential, so with that analogy we can guess $\sigma^{(i)}_\uparrow= -\sigma^*_0$ and $\sigma^{(i)}_\downarrow = 0$. Problem (ii) is solved with Eq. \eqref{3D_circular_disk_charge_distritution}, in which for charge neutrality after superposing (i) and (ii) we need:
\begin{equation}
\begin{split}
& Q^{(ii)} = -(\sigma^{(i)}_\uparrow + \sigma^{(i)}_\downarrow)\pi R^2 \\
\Rightarrow & \  \sigma^{(ii)}_\uparrow(0) = \sigma^{(ii)}_\downarrow(0) = \frac{Q^{(ii)}}{4\pi R^2} = + \frac{\sigma_0^*}{4} \ .
\end{split}
\end{equation}
Therefore, the surface charge densities at the central region are:
\begin{equation}
\begin{split}
&\sigma_\uparrow (0) =  \sigma_\uparrow^{(i)} (0) + \sigma_\uparrow^{(ii)} (0) = -\frac{3\sigma_0^*}4 = -0.75 \sigma^*_0 \ ,
\\
&\sigma_\downarrow (0) =  \sigma_\downarrow^{(i)} (0) + \sigma_\downarrow^{(ii)} (0) = +\frac{\sigma_0^*}{4} = +0.25 \sigma^*_0 \ ,
\end{split}
\end{equation}
clearly different from the results of Eq. \eqref{3D_wrong_solution}. The edge effect has negated the naive infinity assumption with corrections at the same order of magnitude! It should be noted that the sum $\sigma_\uparrow(0)+\sigma_\downarrow(0) \neq 0$ is an indication of strong edge effect, as we can interpret that the charge get pushed away from the center and concentrated at the edge. On the other hand, the difference between them is $\sigma_\uparrow(0) - \sigma_\downarrow(0) = -\sigma_0^*$, since the fringe field is far away (contribution can be ignored) and this relation has to hold so that there is no electrical field inside the conductor.

\ \ 

\section{An Analytical Exploration}

\ \ 

 Some readers might find the above reasoning too hand-waving and need to see a more analytic argument, i.e. from solving the Poisson's equation \cite{poisson1826memoire} directly. This is a difficult task in three-dimensional space, but in two-dimensional space it can be done thanks to the method of conformal mapping which was introduced by J.C. Maxwell \cite{burrow1946application,maxwell1873treatise} and based on the holomorphic transformation of the complexified two-dimensional space. 
 
 \ \ 
 
 We will now show some analytical results in two-dimensional space. The edge effect can already been seen from here. Note that we only choose two-dimensional space out of convenience, and similar physics does happen in three-dimensional space.
 
 \ \ 
 
 \begin{figure*}[!htb]
\centering
\includegraphics[width=0.90\textwidth]{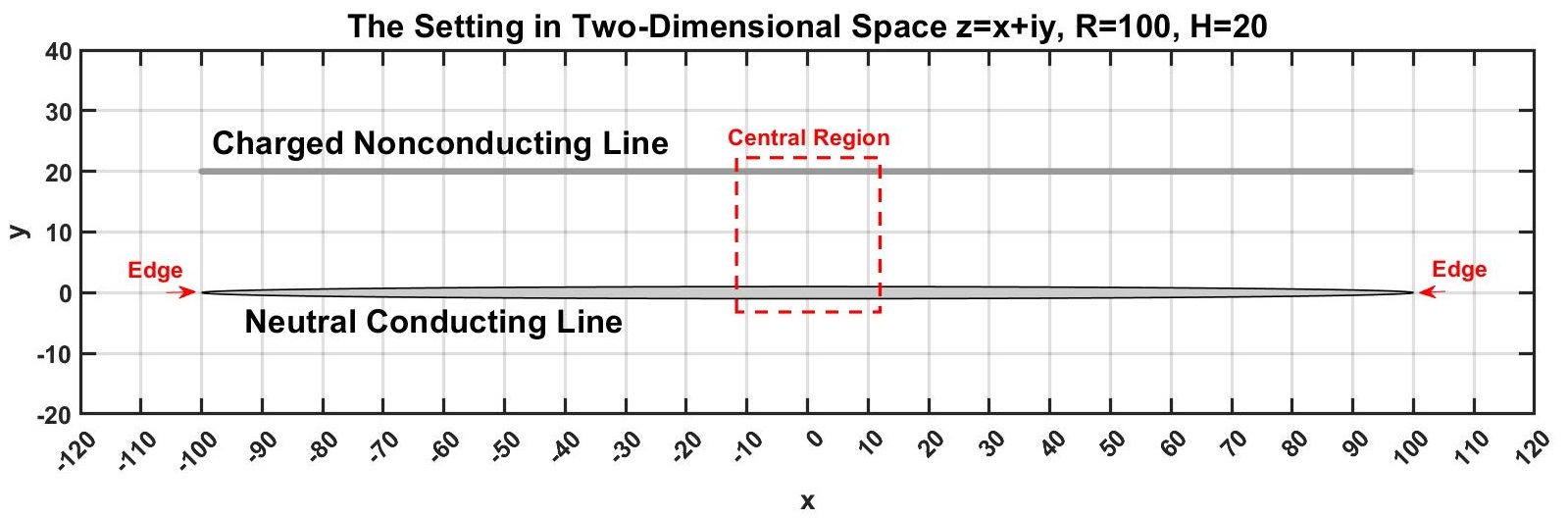}
\caption{A physical setting for the problem of interests in two-dimensional space, with $R=100$ and $H=20$. We also define the central regions to be $|x|\ll R$ and the edges $|x| \rightarrow R$.}
\label{fig:2D_Setting}
\end{figure*}
 
 \ \ 
 
 Choose a Cartesian O$xy$ coordinates, which can be complexified with $z=x+iy$ (where $i$ is the imaginary unit-number). Consider an ellipse conducting region which boundary $(x,y)=(X,Y)$ satisfies:
 \begin{equation} 
 \frac{X^2}{R^2} + Y^2 = 1 \ .
 \label{ellipse_eq}
 \end{equation}
 When $R \gg 1$ and $R \rightarrow \infty$, the ellipse becomes a line of thickness $2$ and total length $2R$. This way of taking limit is analogous to how an oblate ellipsoid becomes a circular disk. At height $y=H>1$, there is a line of charge distribution $\sigma_0(x_c)$ at the horizontal position $x=x_c$ on the line. See Fig. \ref{fig:2D_Setting}.
 
 \ \ 
 
 There is a special Kutta-Joukowski transformation \cite{rektorys2013survey} that take an unit-circle $(u,v)$ in the complexified $w=u+iv$ space and map it to the ellipse $(x,y)=(X,Y)$ given by Eq. \eqref{ellipse_eq} in $z$ space:
\begin{equation}
z = \left( \frac{R+1}2 \right) w + \left( \frac{R-1}2 \right) \frac1w \ .
\label{z_from_w}
\end{equation}
The inverse transformation is as followed:
\begin{equation}
w(x,y) = \frac{\big[x+x_w(x,y)] + i\big[ y+ y_w(x,y) \big]}{R+1} \ ,
\label{w_from_xwyw}
\end{equation}
in which we have to define three more functions prior:
\begin{equation}
\begin{split}
&\Xi(x,y) = R^2 + y^2 - x^2 - 1 \ ,
\\
&x_w(x,y)  = \frac{\sqrt{2} x y }{\sqrt{\Xi(x,y) + \sqrt{4x^2 y^2 + \Xi^2(x,y)}}} \ ,
\\
&y_w(x,y) =
\frac{\sqrt{\Xi(x,y) + \sqrt{4x^2 y^2 + \Xi^2(x,y)}}}{\sqrt{2}} \ .
\end{split}
\label{xwywXi}
\end{equation} 
For a sanity check, we can calculate that $|w(X,Y)|$ always equal to $1$.

 \ \ 
 
 Solving the two-dimensional Poisson's equation for the outside region gives us the complixified potential $\tilde{V}(z)$, in which the real-part is the same as the electrical potential $V(x,y) = \Re [ \tilde{V}(z) ]$. The electrical field $\vec{E} = (E_x,E_y)$ can be calculated from the real-part and imaginary-part of $\partial_z \tilde{V}(z)$: 
\begin{equation}
E_x = -\Re \left[ \partial_z \tilde{V}(z) \right] \ , \ E_y = + \Im \left[ \partial_z \tilde{V}(z) \right] \ .
\label{electric_field}
\end{equation}
To obtain the surface charge density on the center of both sides of the conducting line, we evaluate the electrical field at $X=0$ and $Y=\pm 1$:
\begin{equation}
\sigma(0,\pm 1) = \pm \epsilon_0 E_y = \pm \epsilon_0 \Im \left[ \partial_z \tilde{V}(z) \Big|_{z = \pm i} \right] \ ,
\end{equation}
in which $\sigma_\uparrow(0) = \sigma(0,+1)$ and $\sigma_\downarrow(0) = \sigma(0,-1)$. 

\ \ 

After some exhausting calculation, we obtain the integral form of $\sigma_\uparrow(0)$ and $\sigma_\downarrow(0)$, for a general charge distribution $\sigma_0(x_c)$, conducting line's length $2R$ and separation between lines $H$:
\begin{equation}
\begin{split}
\sigma_\uparrow(0) = \Im \Bigg[ &\int \frac{dx_c \sigma_0(x_c)}{2\pi R} \Bigg( - \frac1{i}  
\\
& \ \ \ \ - \frac1{i - w(x_c,H)} + \frac1{i - \frac1{w^{\dagger}(x_c,H)}}\Bigg) \Bigg] \ ,
\end{split}
\label{2D_sigma_up}
\end{equation}
\begin{equation}
\begin{split}
\sigma_\downarrow (0) = & \left( \frac{R-1}{R+1} \right) \Im \Bigg[ \int \frac{dx_c \sigma_0(x_c)}{2\pi R} \Bigg( - \frac1{i}  
\\
& \ \ \ \ \ \ \ \ - \frac1{i + w(x_c,H)} + \frac1{i + \frac1{w^{\dagger}(x_c,H)}}\Bigg) \Bigg] \ ,
\end{split}
\label{2D_sigma_down}
\end{equation}
where $^\dagger$ is taking the complex conjugation.

\ \ 

For $\sigma_0(x_c)=\sigma_0^*$ in $|x_c| \leq R$ and $0$-value elsewhere, in the limit $R\gg H \gg 1$ and $R \rightarrow \infty$, we can evaluate Eq. \eqref{2D_sigma_up} and Eq. \eqref{2D_sigma_down} to get:
\begin{equation}
\begin{split}
&\sigma_\uparrow (0) = -\frac{(\pi - 1) \sigma^*_0}{\pi} \approx - 0.682 \sigma^*_0 \ ,
\\
&\sigma_\uparrow (0) = + \frac{\sigma^*_0}{\pi} \approx +0.318 \sigma^*_0 \ .
\end{split}
\end{equation}
We show how to get these approximation in Appendix \ref{appendix_integrals}. Similar to the case in three-dimensional space, but with an rigorous and analytical explanation, the sum $\sigma_\uparrow(0) + \sigma_\downarrow (0)$ is non-zero and therefore we can interpret that the edge effect is indeed in control! The difference $\sigma_\uparrow(0) - \sigma_\downarrow (0) = -\sigma_0^*$ is as expected, so no electrical field inside the conductor region.

\ \ 

\section{What Have We Learned?}

\ \ 

When something sounds reasonable, it does not mean it should be correct. That statement is especially true with electrostatics, which is difficult and can be very counter-intuitive due to the lack of daily life's observations and measurements. What we have shown in this paper is a classical version of the phenomena well-known in modern theoretical physics under the name of IR/UV mixing \cite{craig2020ir}, in which the dynamics far away or long time or at low-energy scale ago can shape the local observations at high-energy, high-resolution (and also the other way around, when small disturbance can cause a large emergence behavior). Electrostatics have it, in the form of how the edge effect that happens at the boundary $r\rightarrow R$ can strongly influence the charge distribution deep inside the bulk $r=0$ and the infinity assumption can fail, and unfortunately such phenomena can appear in a relatively simple and familiar setting can be found in many homework assignments of introductory physics courses.

\ \ 

We hope that this notes, at the very least, will stop one of those problems from propagating with an incorrect solution. The correct solution, we believe, is much more advanced, richer in physics and can be a good ``cautionary tale'' for the non-trivialities of electrostatics.

\ \ 

\section{Acknowledgement}

\ \ 

We thank Robert H. Austin for raising this problem and giving useful feedback for the completion of this notes. We thank our colleagues in Princeton University and Jose Gaite for many lively discussions. We thank Long T. Nguyen, Tung X. Tran, Duy V. Nguyen and the xPhO journal club for their support to share this finding to a wider audience. 

\appendix 

\section{Evaluating the Integrals Eq. \eqref{2D_sigma_up} and Eq. \eqref{2D_sigma_down} in some Special Cases \label{appendix_integrals}}

\ \ 

We will consider the cases $R \gg H \gg 1$ and $R \rightarrow \infty$. 

\ \ 

For the case in which the charge is concentrated at $x_c = 0$, say, $\sigma_0(x_c) = +Q^*_0 \delta(x_c)$, then from Eq. \eqref{2D_sigma_up} and Eq. \eqref{2D_sigma_up} we obtain:
\begin{equation}
\begin{split}
\sigma_\uparrow (0) & \approx +\frac{Q_0^*}{2\pi R} - \frac{Q_0^*}{2 \pi H} \left( 1 + \sqrt{1 + \frac{H^2}{R^2}} \right)
\\
& \approx + \frac{Q_0^*}{2\pi R} -\frac{Q_0^*}{\pi H} + \mathcal{O}(H) \rightarrow -\frac{Q_0^*}{\pi H} \ ,
\end{split}
\end{equation}
\begin{equation}
\begin{split}
\sigma_\downarrow (0) &\approx +\frac{Q_0^*}{2\pi R} + \frac{Q_0^*}{2\pi H} \left( 1 - \sqrt{1 + \frac{H^2}{R^2}}\right) 
\\
& \approx +\frac{Q_0^*}{2\pi R} + \mathcal{O}(H) \rightarrow 0 \ .
\end{split} 
\end{equation}
This distribution of surface charge on the conducting line in the central region is the same with the situation in which it is grounded. The correction due to charge neutrality condition and the edge effect is there, but weak and negligible (unless we are looking very close to the two edges at $x=\pm R$).

\ \ 

\begin{figure}[!htb]
\centering
\includegraphics[width=0.45\textwidth]{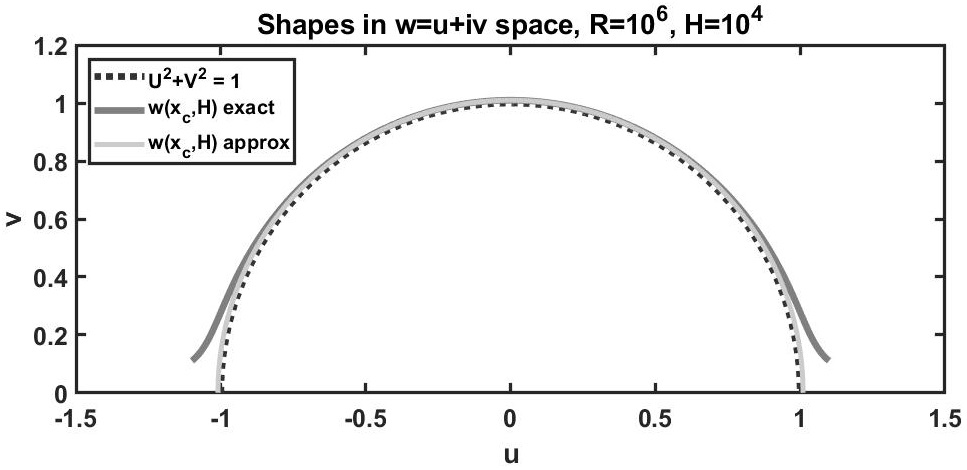}
\caption{Comparison between $W(x_c,H)$ approx and $w(x_c,H)$ exact for $R=10^6$ and $H=10^4$.}
\label{fig:wexact_and_wapprox}
\end{figure}

For the case in which the charge distribution is uniform $\sigma_0(x_c) = \sigma_0^*$ in $x_c \in  [-R,+R]$ (and $0$-value elsewhere), a good approximation for this setting is to prioritize the intermediate central region $R\gg |x_c| \gg H \gg 1$ and keep up to the terms $\mathcal{O}\left(Hx_c \right)$:
\begin{equation}
x_w(x_c,H) \approx \frac{Hx_c}{R} \ , 
 \ y_w(x_c,H) \approx R - \frac{x_c^2}{2R} \ .
\end{equation}
With these formulas, we can approximate:
\begin{equation}
\begin{split}
w(x_c,H) & \approx \left( 1 + \frac{H}{R} \right) \frac{x_c}{R} + i \left( 1 + \frac{H}{R} - \frac{x_c^2}{2R^2} \right) 
\\
&  \approx  \hat{x}_c + i \sqrt{\hat{1}^2 - \hat{x}_c^2}  \  ,
\end{split}
\end{equation}
in which we check that the second-line can be series-expanded to give the first-line if we choose: 
\begin{equation}
\hat{x}_c = \left( 1 + \frac{H}{R}\right) \frac{x_c}{R} \ , \ \hat{1} = 1+\frac{H}{R}  \ .
\end{equation}
We can also check with numerical calculation \cite{matlab2021mathworks} that this $w(x_c,H)$ approximation is in good agreement with $w(x_c,H)$ exact, as shown in Fig. \ref{fig:wexact_and_wapprox}.

\ \ 

\ \

Eq. \eqref{2D_sigma_up} and Eq. \eqref{2D_sigma_down} can then be evaluated:
\begin{equation}
\begin{split}
&\sigma_\uparrow (0) \approx  + \frac{\sigma_0^*}{\pi} 
\\
& \ \ \ \ - \frac{\sigma_0^*}{2\pi \hat{1}}  \int^{+\hat{1}}_{-\hat{1}} d\hat{x}_c \left[ \frac{  \hat{1}^2 - 1 }{ \hat{1}^2 + 1  - 2\sqrt{\hat{1}^2 - \hat{x}_c^2} } \right] \\
& \approx + \frac{\sigma_0^*}{\pi}  - \frac{\sigma_0^*}{2\pi \hat{1}} \Big[ 2\pi + \mathcal{O}(H) \Big] 
 \approx -\left(1-\frac1\pi \right) \sigma_0^* \ ,
\end{split}
\label{2D_sigma_up_approx}
\end{equation}
\begin{equation}
\begin{split}
&\sigma_\downarrow (0) \approx + \frac{\sigma_0^*}{\pi} 
\\
& \ \ \ \ - \frac{\sigma_0^*}{2\pi \hat{1}}  \int^{+\hat{1}}_{-\hat{1}} d\hat{x}_c \left[ \frac{  \hat{1}^2 - 1 }{ \hat{1}^2 + 1  + 2\sqrt{\hat{1}^2 - \hat{x}_c^2} } \right] 
\\
& \ \ \approx + \frac{\sigma_0^*}{\pi}  - \frac{\sigma_0^*}{2\pi \hat{1}} \Big[ 0 + \mathcal{O}\left( H\right) \Big] \approx + \frac{\sigma_0^*}{\pi} \ .
\end{split}
\label{2D_sigma_down_approx}
\end{equation}
Note that, for the last lines of Eq. \eqref{2D_sigma_up_approx} and Eq. \eqref{2D_sigma_up_approx}, we use the integral:
\begin{equation}
\begin{split}
\int d\hat{x}_c &\left[ \frac{\hat{1}^2-1}{\hat{1}^2+1 \mp 2\sqrt{\hat{1}^2 - \hat{x}_c}} \right]
\\
& = \frac{\hat{1}^2 + 1}2 \arctan \left[ \frac{2\hat{x}_c}{\hat{1}^2 -1} \right]
\\
& \ \ \ \mp \frac{\hat{1}^2 + 1}2 \arctan \left[ \frac{(\hat{1}^2-1)\sqrt{\hat{1}^2 - \hat{x}_c^2}}{(\hat{1}^2+1)\hat{x}_c} \right]
\\
& \ \ \ \mp \frac{\hat{1}^2 -1}{2} \arcsin \left[ \frac{\hat{x}_c}{\hat{1}} \right] \ .
\end{split}
\end{equation}

\ \ 

\bibliography{main}
\bibliographystyle{apsrev4-2}

\end{document}